\documentclass[]{spie}  %>>> use for US letter paper

\pdfoutput=1

\usepackage{amsmath,amsfonts,amssymb}
\usepackage{graphicx}
\usepackage[colorlinks=true, allcolors=blue]{hyperref}
\usepackage{wrapfig}

\usepackage{placeins}
\usepackage{caption}
\usepackage{subcaption}
\usepackage{graphbox}
\usepackage[export]{adjustbox}
\usepackage{gensymb}

\title{Physical Modeling and Performance of Spatial-Spectral Filters for CT Material Decomposition}
\author[]{\vspace{-3mm}Matthew Tivnan, Steven Tilley II, J. Webster Stayman}
\affil[]{\normalsize\vspace{-3mm}\em Department of Biomedical Engineering, Johns Hopkins University, Baltimore MD, USA 21205}

\begin{document} 
\maketitle
\vspace{-2em}
\begin{abstract}
\vspace{-0.5em}
Material decomposition for imaging multiple contrast agents in a single acquisition has been made possible by spectral CT: a modality which incorporates multiple photon energy spectral sensitivities into a single data collection. This work presents an investigation of a new approach to spectral CT which does not rely on energy-discriminating detectors or multiple x-ray sources. Instead, a tiled pattern of K-edge filters are placed in front of the x-ray to create spatially encoded spectra data. For improved sampling, the spatial-spectral filter is moved continuously with respect to the source. A model-based material decomposition algorithm is adopted to directly reconstruct multiple material densities from projection data that is sparse in each spectral channel. Physical effects associated with the x-ray focal spot size and motion blur for the moving filter are expected to impact overall performance. In this work, those physical effects are modeled and a performance analysis is conducted. Specifically, experiments are presented with simulated focal spot widths between 0.2~mm and 4.0~mm. Additionally, filter motion blur is simulated for a linear translation speeds between 50~mm/s and 450~mm/s. The performance differential between a 0.2~mm and a 1.0~mm focal spot is less than 15\% suggesting feasibility of the approach with realistic x-ray tubes. Moreover, for reasonable filter actuation speeds, higher speeds are shown to decrease error (due to improved sampling) despite motion-based spectral blur.
\end{abstract}

%\keywords{Manuscript format, template, SPIE Proceedings, LaTeX}

\vspace{-1em}
\section{INTRODUCTION}
\label{sec:intro}  % \label{} allows reference to this section
\vspace{-0.5em}
Multi-contrast agent imaging is an active area of research. For example, iodine and gadolinium have been used together in various applications including multi-phase kidney and liver imaging \cite{symons2017photon}, colonography \cite{muenzel2016spectral}, and post-operational imaging for endovascular aneurysm repair \cite{dangelmaier2018experimental} among others. Iodine, gold, and calcium phosphate have also been used as target materials to study atherosclerotic plaque composition \cite{cormode2010atherosclerotic,baturin2012spectral}. New technology which allows for decomposition into more material components or improved low-concentration estimation will greatly benefit the field of multi-contrast imaging.

\begin{wrapfigure}[13]{r}{0.44\textwidth}
\centering
	\vspace{-10mm}
  \includegraphics[width=0.4\textwidth]{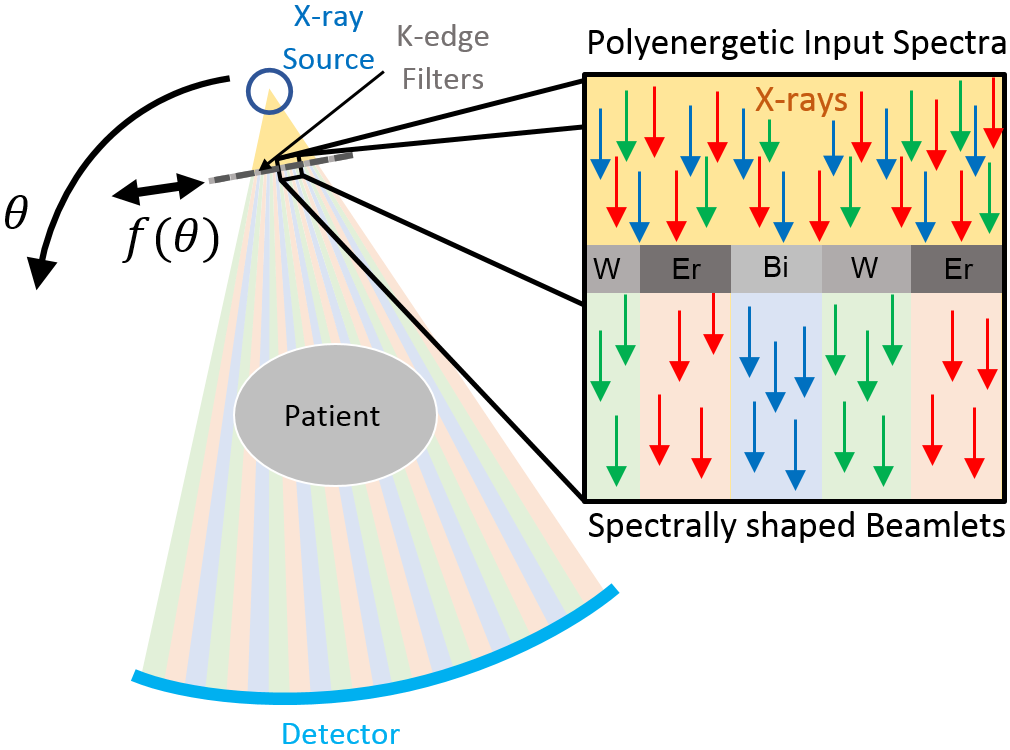}
\caption{Spectral CT using moving spatial-spectral filters and energy-integrating detectors.}
\label{fig:spectralCT}
\end{wrapfigure}

Developments have focused on incorporating different and varied spectral sensitivities into measurements to enable \emph{spectral CT}. Methods include dual sources \cite{flohr2006first}, kV-switching \cite{xu2009dual}, split filters \cite{rutt1980split}, dual-layer-detectors \cite{carmi2005material}, and photon-counting detectors \cite{schlomka2008experimental}. With the exception of photon-counting, these methods typically offer only two spectral channels.

A new method to enable spectral CT with ordinary energy-integrating detectors is shown in Figure~\ref{fig:spectralCT}.
Specifically, a ``spatial-spectral'' filter, composed of a repeating pattern of K-edge filter materials, is placed in front of the x-ray source dividing the full x-ray beam into spectrally varied beamlets \cite{stayman2018model}. 
The filter is translated parallel to the detector as the CT gantry rotates to provide spatially interlaced projection data with different spectral channels. Since each spectral channel is sparse, conventional reconstruction methods involving material decomposition in the projection domain or the image domain are ill-suited for data processing. In contrast, a model-based material decomposition (MBMD) algorithm \cite{tilley2018model} permits simultaneous processing of all data as well as sophisticated regularization schemes (e.g. compressed sensing) to overcome traditional sampling limitations. Advantages of the  spatial-spectral filter include flexibility in spectral shaping, scaling to include more spectral channels, and possible low-cost integration into current CT systems. There is also the potential to combine this approach with other approaches to extend low-concentration performance.

Preliminary investigations \cite{stayman2018model} demonstrated the feasibility of the spatial-spectral filtering approach under highly idealized conditions. In this work, a more accurate physical model is developed, taking into account focal spot effects and motion blur associated with the moving filter. This more accurate model is used to investigate potential performance limitations and to guide future spatial-spectral CT system design. Performance is evaluated in a multi-contrast digital CT phantom across a range of practical focal spot sizes and filter speeds.

%\begin{itemize}
%	\item Growing demand from clinical applications with multiple contrast agents
%	\item Spectral CT is the technology which will allow for sufficient material discrimination and high spatial resolution for multi-contrast-enhanced imaging
 %   \item Spectral CT involves acquiring lines of projection with different and varied spectral responses
%    \item Existing approaches include dual sources, KV switching, split filter, dual-layer detectors, photon counting detectors. 
%    \item Advantages of spatial-spectral filter: 
%    \begin{itemize}
%    	\item Flexibility of materials and patterns: can be optimized for contrast agent types
%        \item Scales easily to more and more spectral channels
%        \item MBIR available for direct reconstruction: fewer artifacts (?)
%    \end{itemize}
%    \item Previous proposal used an ideal model without blur.
%    \item Blur affects not only spatial resolution, as in traditional CT, but also the spatial-spectral distribution
%    \item This work intends to model focal spot blur and filter motion blur
%    \item Simulations presented here will be used to answer questions about effects of blur on data acquisition and image reconstruction with MBIR. 
%    \item Results of these numerical experiments will also be used to inform the design of a physical spatial-spectral filter
%\end{itemize}

\FloatBarrier
\vspace{-1em}
\section{METHODS}

\subsection{General Physical Model for CT Acquisitions with Spatial-Spectral Filters}

A general forward model for spectral CT with varied spectral sensitivities and energy-integrating detectors is

%\begin{equation}
%	y(u,\theta) = \int_E S(u,\theta, E) \exp{\Big(-\sum_j l_j(u,\theta)  \mu_j(E) \Big)} dE \quad \quad , \quad \quad \mu_j(E) = \sum_k  x_{j,k} \enspace q_k(E) 
%\end{equation}

\setlength{\abovedisplayskip}{0pt}
\setlength{\belowdisplayskip}{3pt}
\setlength{\abovedisplayshortskip}{0pt}
\setlength{\belowdisplayshortskip}{3pt}
\vspace{-3mm}
\begin{equation}
\vspace{-3mm}
	y_i = y(u_i,\theta_i) = \int_E S(u_i,\theta_i,E) \exp{\Big(-\sum_j l_j(u_i,\theta_i)  \mu_j(E) \Big)} dE, \quad \quad \quad \mu_j(E) = \sum_k  {\rho}_{j,k} q_k(E) 
\end{equation}

\noindent where $y_i$ is the $i^{th}$ measurement, a sample of the projection data, $y(u_i,\theta_i)$, at detector position, $u_i$, and rotation angle, $\theta_i$. The system spectral response, $S$, is measurement-dependent, $l_j$ are projection contributions of the $j^{th}$ voxel, and $\mu_j(E)$ is the energy-dependent attenuation coefficient of the $j^{th}$  voxel. The latter coefficient is modeled as a weighted sum over material index $k$ of material basis functions $q_k(E)$ weighted by material densities, $\rho_{j,k}$ for each voxel. This model is extremely general. For example, for a kV-switching CT system, $S(u,\theta, E)$ is equal to a high-kV spectrum, $s_H(E)$ for $\theta_i$ with odd indexes and to a low-kV spectrum $s_L(E)$ for $\theta_i$ with even views.
One may define $S$ for an ideal spatial-spectral CT system as $S(u,\theta,E) = S_0(u+f(\theta),E)$ where $S_0(u,E)$ represents a spatial function of all beamlet energies across the detector (found, e.g., by computing the polyenergetic spectrum that exits each sub-filter). The filter is translated laterally according to the function $f(\theta)$ with rotation angle. Unfortunately this model excludes some important physical aspects of a real spatial-spectral system.

\begin{figure}[b!]
\centering\vspace{4mm}
\begin{subfigure}[t]{.23\textwidth}
  \centering
  \includegraphics[trim = {0 8.5mm 30mm 6mm},clip,align=c,height=38mm, keepaspectratio]{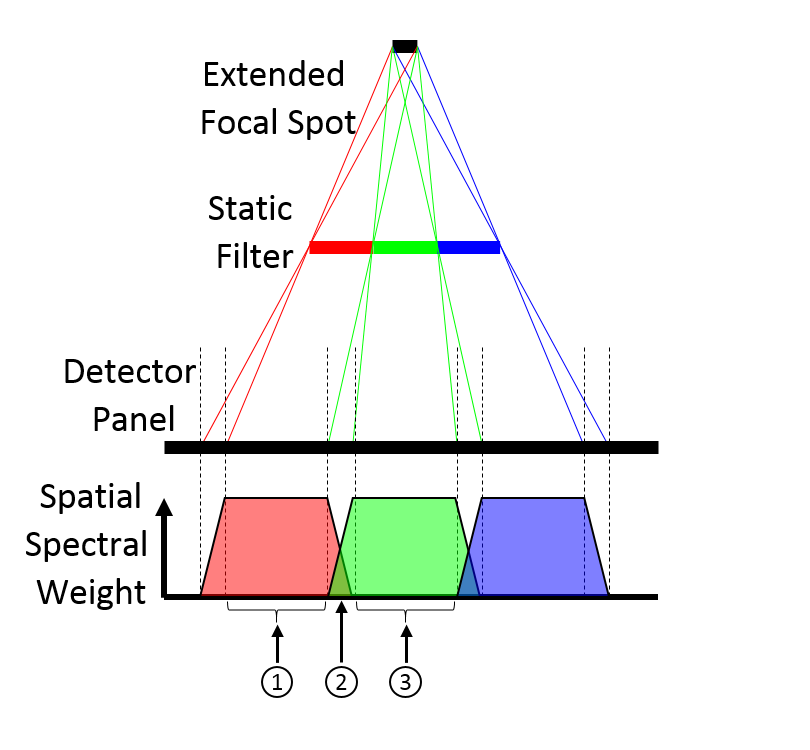}
  \caption{}
  \label{fig:focalSpotBlur}
\end{subfigure}%
\begin{subfigure}[t]{.23\textwidth}
  \centering
  \includegraphics[trim = {6mm 9mm 0 8mm},clip,align=c,height=38mm, keepaspectratio]{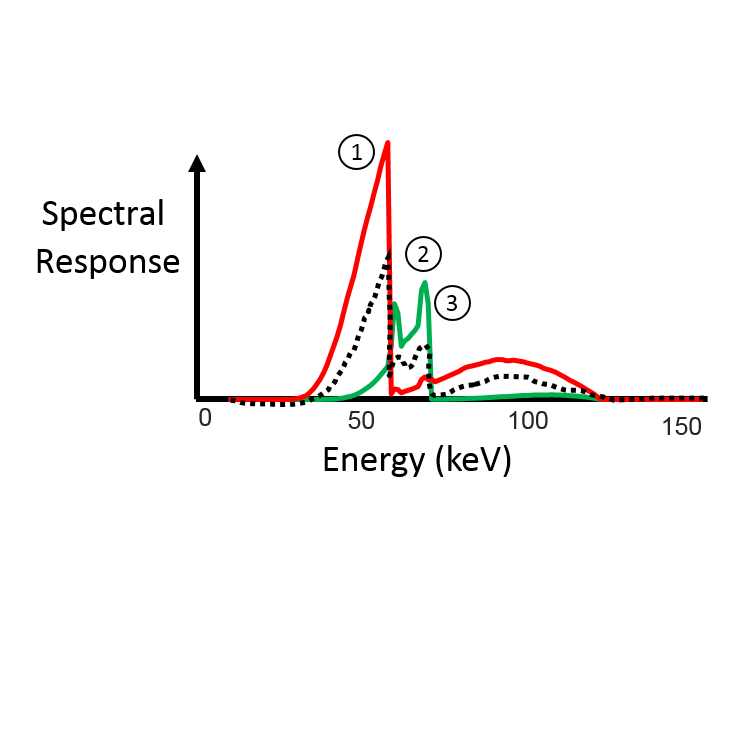}
  \caption{}
  \label{fig:spectra}
\end{subfigure}%
\begin{subfigure}[t]{.23\textwidth}
  \centering
  \includegraphics[trim = {0 9mm 30mm 6mm},clip,align=c,height=38mm, keepaspectratio]{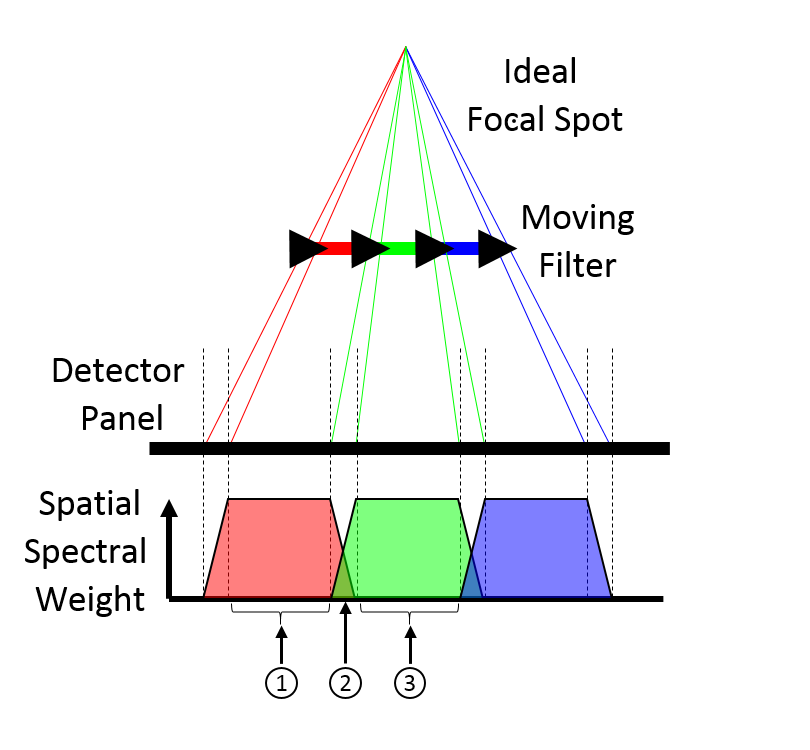}
  \caption{}
  \label{fig:filterMotionBlur}
\end{subfigure}
\begin{subfigure}[t]{.15\textwidth}
  \centering
  \includegraphics[trim = {10mm 9mm 10mm 0},clip,align=c,height=38mm, keepaspectratio]{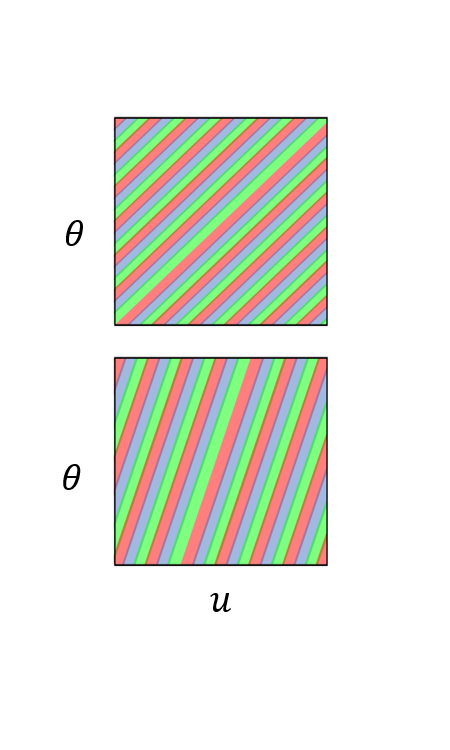}
  \caption{}
  \label{fig:spectralSampling}
\end{subfigure}
\caption{Focal spot (a), filter motion (c), spectral blur (b), and the spectral response, $S(u,\theta,E)$ (d).}
\label{fig:spectralBlur}
\end{figure}

First, realistic x-ray focal spots are extended resulting in blur of objects in the imaging system. While such blur is relatively minor for objects at the center of the field-of-view, filters placed near the x-ray source will ``see'' significant blur due to magnification effects. This has the effect of mixing the spectra of neighboring beamlets as illustrated in Figure~\ref{fig:focalSpotBlur} and \ref{fig:spectra}. For a thin filter and a flat detector, this blur is accurately modeled by a convolution applied to the ideal spectrum $S_0$. We approximate the shape of a realistic focal spot distribution using a rectangular kernel, $h_{FS}(u)$, with width equal to the focal spot width magnified by the ratio between the filter-to-detector distance and the source-to-filter distance.
The second important effect involves the moving filter. In particular, for realis andtic CT gantry rotation rates, step-and-shoot motion of the filter is impractical. We consider the more realistic case where the filter moves continuously including during the detector integration interval. For a fixed gantry rotational speed, the spatial-spectral sampling profile is defined by $f(\theta) = \alpha \enspace\theta$, where $\alpha$ is proportional to filter speed. This motion changes both the spatial-spectral sampling (Figure~\ref{fig:spectralSampling}) and imparts an additional spatial blur of spectra (Figure~\ref{fig:filterMotionBlur}) which we model by a convolution with a second kernel $h_M(u)$. We module this kernel as rectangular with width equal to the distance the filter moves per view magnified by the ratio of the source-to-detector distance and the source-to-filter distance. 

%\begin{figure}[b!]
%\centering
%\includegraphics[align=c,width=.8\textwidth]{figures/spectral_blur}
%\caption{Spectral blur effects due to an extended x-ray focal spot (left) and filter motion (right). Spatial-spectral sampling for fast (top-right) and slow (bottom-right) filter speeds}
%\label{fig:spectralBlur}
%\end{figure}

Thus, the overall spectral model with these physical effects modeled may be written as
\begin{equation}
\vspace{-3mm}
	S(u,\theta,E) = h_M(u) * h_{FS}(u) * S_0 (u + f(\theta),E) 
\vspace{2mm}
\end{equation}
Any filter motion pattern can be chosen and modeled by $f(\theta)$, but for the remainder of this work we will consider the constant-speed linear filter motion given by $f(\theta) = \alpha \enspace\theta$. The compounding effect of the operators $h_M(u) * (\cdot{})$ and $h_{FS}(u) * (\cdot{})$ will be a blur of spectra in the projection domain. 
Although the extended focal spot and filter motion can be characterized separately, both effects occur together in physical acquisitions. The experimental methods presented below include both filter motion effects and focal spot blur despite the diagram in Figure \ref{fig:spectralBlur} which shows the two effects separately.

\vspace{-2em}
\subsection{Simulation Study on Spatial-Spectral Performance}
\vspace{-0.5em}
In general, one would expect that the mixed spectral responses will degrade the ability to separate different materials even when those spectral are appropriately modeled. The overall impact of filter speed on material decomposition performance is potentially more complex. If the filter speed is zero, the spatial-spectral sampling is poor. (E.g., For a static filter, the central detector measurement will only be probed by a single spectral channel.) However, as the filter speed increases, the motion blur effects are more dramatic. For extremely fast motion, all spectral channel can potentially blur together. Between these two extremes and within the constraints of realistic filter speeds there may be an optimum filter speed. 
Numerical experiments were employed to characterize the impact of these physical effects on MBMD estimation performance.

\begin{wraptable}[8]{r}{0.3\textwidth}
\vspace{-3mm}
\centering\scriptsize 
\begin{tabular}{|l|l|}
\hline
 source-filter distance & 380~mm \\ \hline
 source-isocenter distance & 890~mm \\ \hline
 source-detector distance & 1040~mm  \\ \hline
 gantry rotation speed & 120~RPM  \\ \hline 
 views per rotation & 360  \\ \hline
 projections per view & 512 \\ \hline
 pixel size & 0.556~mm  \\ \hline
 image space dimensions & 128~$\times$~128  \\ \hline
 voxel size (square) & 0.5~mm  \\ \hline
\end{tabular}
\caption{Geometry and sampling.}
\label{tab:CTgeom}
\end{wraptable}
\normalsize

\begin{figure}
	\centering
	\includegraphics[width=0.8\linewidth]{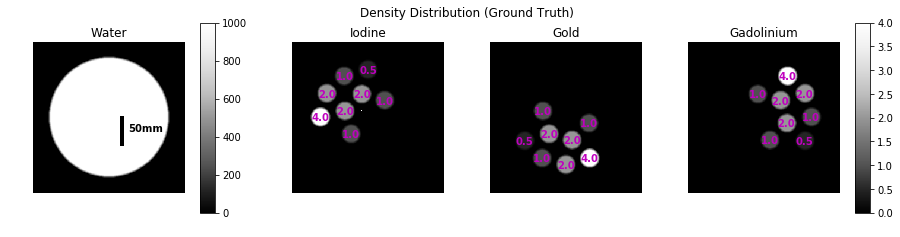}
  	\label{fig:images_ground_truth}
	\caption{Ground truth of the numerical phantom. Magenta text indicates the density in mg/mL of iodine, gold, or gadolinium (corresponding to image subtitle) in cylindrical inserts.}
\label{fig:numericalPhantom}
\vspace{-1.3em}
\end{figure}

The geometry and sampling conditions for the studies are summarized in Table \ref{tab:CTgeom}. A digital phantom (Figure~\ref{fig:numericalPhantom}) of a 100~mm diameter water cylinder and several 15mm diameter cylindrical inserts containing various mixtures of iodine, gold, and gadolinium was employed. The outer ring includes single-contrast inserts of 0.5-4.0~mg/mL concentrations. The inner ring includes mixtures of 1.0~mg/mL and 2.0~mg/mL for all combinations of two materials.  The center of the phantom also includes 10.0~mg/mL single voxel impulses of each material for regularization tuning.

The filter materials were chosen based on a previous study \cite{stayman2018model} that tested all three and four filter-material combinations to maximize multi-contrast-agent concentration estimation performance. Specifically, we select a filter comprised of 0.25~mm-thick, 1.46~mm-wide strips of bismuth, gold, lutetium, and erbium. Thus, each spectral beamlet covers an area on the detector that is 8~pixels wide.  Incident fluence was uniform across the filter and the level was adjusted such that the bare-beam fluence for the bismuth-filtered beamlet was $10^5$ photons/pixel.

In the first numerical experiment, we simulated focal spot widths of 0.2-4.0~mm and held filter motion speed constant at 131.4~mm/s which corresponds to one detector pixel per view after magnification. In the second experiment, we simulated filter motion speeds between 50-450~mm/s and held the focal spot width constant at 0.4~mm.

We used the MBMD algorithm with the new models for focal spot blur and filter motion effects to reconstruct density distributions for the four materials present in the phantom. All numerical experiments used 1000 iterations of the algorithm. We used a quadratic regularizer with material-dependent regularization strengths which were tuned such that the FWHM of the PSF corresponding to the 10.0~mg/mL voxel impulse was 1.8~mm~$\pm$~0.2~mm for all target materials, focal spot widths, and filter motion speeds. Importantly, our current aim is not to analyze the impact of a mis-match between the reconstruction model used by the MBMD algorithm and the true acquisition parameters. Rather, we implement a matched reconstruction model and aim to characterize the image degradation when transition between spectra is blurred.

Root-mean-squared error (RMSE) was used for analysis and was computed by first finding the RMSE within regions of interest (ROIs) inside each cylindrical insert and then taking the mean across all ROIs. This was done separately for each target material.

\FloatBarrier

\section{RESULTS}

% START with the imaging results for a 1~mm focal spot... That is a fairly standard size. Discuss that that the decomposition basically works even with focal spot blur...

In the imaging results for the focal spot experiment, the 0.2~mm and 1.0~mm focal spot width cases are very difficult to distinguish by eye. Overall the reconstructed densities are a reasonable approximation of the ground truth. The low-contrast 0.5~mg/mL insert is visible in both cases which implies that the spatial-spectral filter has the potential to improve sensitivity to lower concentrations. A focal spot size of 1.0mm is fairly standard, so the material decomposition appears to be effective in the presence of realistic focal spot blur effects. 

\begin{wrapfigure}[13]{r}{.66\textwidth}
\vspace{-5mm}
\centering
\begin{subfigure}{.33\textwidth}
  \centering
  \includegraphics[width=\linewidth]{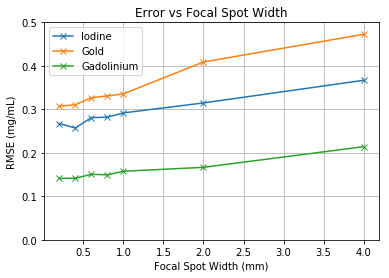}
  \caption{}
  \label{fig:trendFocalSpot}
\end{subfigure}%
\begin{subfigure}{.33\textwidth}
  \centering
  \includegraphics[width=\linewidth]{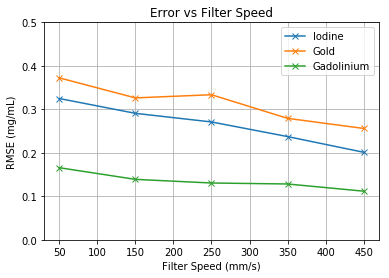}
  \caption{}
  \label{fig:trendFilterSpeed}
\end{subfigure}
\caption{Error vs focal spot blur (a) and filter motion (b).}
\label{fig:trendPlots}
\end{wrapfigure}

%%% WEB - use tilde for a non-breaking space between the number and the units...
For focal spot widths between 0.2-1.0~mm, the final RMSE values were less than 0.35~mg/mL for each material. In the case of gadolinium, the RMSE was less than 0.18~mg/mL for this range. Larger-than-average focal spot widths such as 2.0~mm and 4.0~mm resulted in RMSE values around 0.47~mg/mL. The overall trend shows that larger focal spot widths lead to greater error. However, the change in RMSE is less than 15\% between the 0.2~mm and 1.0~mm cases for any individual contrast agent so the impact is not severe. One notable error is the insert containing 4.0~mg/mL of iodine on the left side of the image. The reconstructed density of iodine is underestimated at 2.75~mg/mL and around 1.25~mg/mL is erroneously attributed to gold. This could indicate that for the given combination of filter materials, iodine and gold are particularly difficult to distinguish. This issue may also be improved with a more sophisticated regularization scheme, more spectral channels, or higher fluence.

In the filter speed experiment, RMSE consistently decreased for higher filter speeds. For all materials, the RMSE of the 450~mm/s filter speed was around 40\% lower than the 50~mm/s case. This result would seem to indicate that in the realistic range of filter motion speeds, the benefits of improved spatial-spectral sampling outweigh the negatives of filter motion spectral blur.

\begin{figure}
	\centering
	\includegraphics[width=0.8\linewidth]{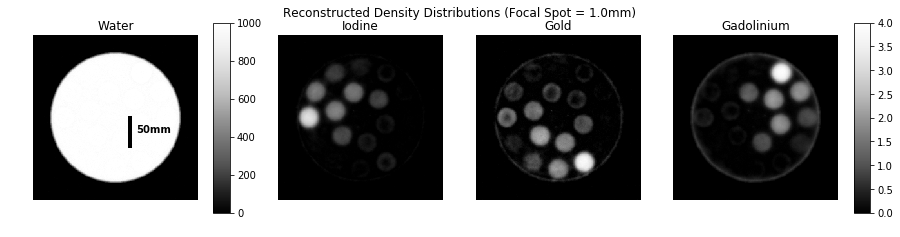}
  	\label{fig:sub1}
	\caption{Example of a material decomposition result for a focal spot width of 1.0mm.}
\label{fig:reconExample}
\end{figure}

%%% I don't think we'll have room for this...
%\begin{figure}
%	\centering
%    \includegraphics[width=\linewidth]{figures/bars_FSW_1000}
%    \caption{Reconstructed densities in each ROI for iodine (red), gold (green), and gadolinium (blue). The colored bars represent the ground truth, and the center markers and error bars represent the reconstructed means and standard deviations for simulation results with focal spot width of 1.0mm. }
%\label{fig:barPlot}
%\end{figure}

\section{Conclusion}

The focal spot blur experiment has shown that error increases as focal spot size increases. However, for realistic focal spot sizes between 0.2~mm and 1.0~mm there is a relatively small change in performance. This suggests that, as long as spectra are modeled for each measurement, spatial-spectral filters are viable for use with a range of realistic x-ray sources. The filter motion experiment demonstrates the importance of the spatial-spectral sampling pattern on the MBMD algorithm's ability to separate various target materials. Spectral blur effects from filter motion were shown to be outweighed by the benefits of improved sampling for the filter speed range in the study. Overall, error decreased as filter speed increased - giving finer spectral sampling over projection angles. Knowledge of this performance trade-off will be valuable for the next stages of this work where choice of filter speed must be balanced with the realistic range of speeds that can be precisely controlled in a CT acquisition. For example linear motor have been investigated for CT filter actuation with speeds up to 5000~mm/s but this is not necessarily achievable with sufficient precision or within acceleration constraints. 

In light of the results presented in this work, it would be prudent to revisit the optimization of filter design with this improved physical model. The order of filter materials may now have a greater impact since the spectral blur occurs between neighboring beamlets. We will also need to characterize the impact of reconstruction model mismatches and develop calibration methods. As we build upon our understanding of this new technology, we move closer to the physical implementation of a spatial-spectral filter system with the ultimate goal of heightening sensitivity to low concentrations and improving material discrimination for multi-contrast-enhanced CT.

%%% Future work thoughts:
% Reconsider design optimization with improved physical model
% Order of materials may be important (since there is spectral mixing)
% Explore mismatches and strategies for calibrating spectra in a real system
% Make some design choices and start physical implementation

%%% The following is OK, I have some ideas closer to the research plan above... I like ending on a strong positive as you've done below. Consider how to do something similar with the above list.
%In the future, spectral blurring and sampling effects from the extended focal spot and filter motion must be given proper consideration in light of these experiments. However, in no way do these effects pose an insurmountable barrier for the technology. In general these experiments affirm that the spatial-spectral filter is useful beyond an ideal CT acquisition model. It appears to be a valid method for spectral CT under realistic conditions. These results are promising for the prospect of the design and construction of a physical spatial-spectral filter in order to move beyond simulation. 

\acknowledgments % equivalent to \section*{ACKNOWLEDGMENTS} 
This work was supported, in part, by NIH grant R21EB026849.

\bibliography{report}{}
\bibliographystyle{spiebib-abbr} % makes bibtex use spiebib.bst

\end{document}